\documentclass[aps,prb,reprint,superscriptaddress,showpacs,showkeys]{revtex4-1}

\usepackage{graphicx}
\usepackage{bm}

\begin{document}


\title{$^{51}$V-NMR study of antiferromagnetic state and spin dynamics \\ in the quasi-one-dimensional BaCo$_2$V$_2$O$_8$}



\author{Yukiichi Ideta}
\affiliation{Department of Quantum Materials Science, Institute of Technology and Science, The University of Tokushima, Tokushima 770-8506, Japan}
\author{Yu Kawasaki}
\email{Corresponding author: yu@pm.tokushima-u.ac.jp}
\affiliation{Department of Quantum Materials Science, Institute of Technology and Science, The University of Tokushima, Tokushima 770-8506, Japan}
\author{Yutaka Kishimoto}
\affiliation{Department of Quantum Materials Science, Institute of Technology and Science, The University of Tokushima, Tokushima 770-8506, Japan}
\author{Takashi Ohno}
\affiliation{Department of Quantum Materials Science, Institute of Technology and Science, The University of Tokushima, Tokushima 770-8506, Japan}
\author{Yoshitaka Michihiro}
\affiliation{Department of Quantum Materials Science, Institute of Technology and Science, The University of Tokushima, Tokushima 770-8506, Japan}
\author{Zhangzhen He}
\affiliation{State Key Laboratory of Structural Chemistry, Fujian Institute of Research on the Structure of Matter, Chinese Academy of Sciences, Fuzhou, Fujian 350002, P.R.\ China}
\author{Yutaka Ueda}
\affiliation{Institute for Solid State Physics, University of Tokyo, Kashiwa 277-8581, Japan}
\author{Mitsuru Itoh}
\affiliation{Materials and Structures Laboratory, Tokyo Institute of Technology, Midori, Yokohama 226-8503, Japan}


\date{\today}

\begin{abstract}
We report on our $^{51}$V-NMR study of static and dynamical magnetic properties in the quasi-one-dimensional antiferromagnet BaCo$_2$V$_2$O$_8$\@. 
Although the NMR spectrum shows well-defined antiferromagnetic (AF) order in the N\'eel ground state, the AF characteristic from the NMR spectrum is incomplete between 3.5 K and $T_N=$ 5.4 K, which could be affected by quantum spin fluctuations. 
The AF NMR spectrum indicates two V sites experiencing different magnetic field magnitudes, $H_{A1}=2.1$ kOe and $H_{A2}=3.8$ kOe.
These internal fields could be explained by accounting for the classical and the pseudo-dipolar fields from Co$^{2+}$ spins with a proposed magnetic structure based on the neutron diffraction measurements. 
In the paramagnetic state, the nuclear spin relaxation is dominated by AF spin fluctuations through the dipolar-type coupling between V and surrounding Co$^{2+}$ ions.
The linear relation between the nuclear spin-lattice relaxation rate $1/T_1T$ and the magnetic susceptibility $\chi$ indicates that the $Q$ component of magnetic susceptibility $\chi(Q)$ is roughly proportional to $\chi$, where $Q$ is the AF wave number.
A change in slope of $1/T_1T$ with respect to $\chi$ around 150 K suggests a change in the AF spin fluctuation spectrum.
\end{abstract}


\maketitle

\section{Introduction}

One-dimensional (1D) quantum spin systems have attracted considerable attention in the past decades, because these display a rich variety of ground states with nonclassical magnetic phenomena.
An ideal 1D antiferromagnetic (AF) spin system is well known not to show long-range order at finite temperatures due to strong quantum spin fluctuations.\cite{bethe31}
Nevertheless, slight perturbations of the system by, for instance, weak inter-chain interactions or doping with nonmagnetic impurities, can destabilize the quantum critical state and favor three-dimensional (3D) AF long-range order.\cite{santos90,uchiyama99}

Recently, the field-induced transition from 3D AF long-range order to 1D quantum disorder was discovered in the quasi-1D AF spin-chain system BaCo$_2$V$_2$O$_8$\@.\cite{he05prb}
This transition was theoretically predicted originally for gapped spin systems doped with nonmagnetic impurities.\cite{mikeska04}
The crystal structure of BaCo$_2$V$_2$O$_8$ (Fig.\ 1) is characterized by edge-shared CoO$_6$ octahedra forming a screw chain along the $c$-axis.
The screw chains are separated by non-magnetic VO$_4$ (V$^{5+}$) tetrahedra and Ba$^{2+}$ ions, resulting in a quasi-1D structural arrangement of magnetic Co$^{2+}$ ions.\cite{he05cm,he06jcg,wichmann86}

In BaCo$_2$V$_2$O$_8$, the orbital angular momentum of Co$^{2+}$ is not completely quenched and the resulting low-energy Kramers doublet is considered as a pseudo spin $S=1/2$\@.\cite{kimura06jp}
The temperature dependence of magnetic susceptibility $\chi$ shows a broad maximum around 30 K signaling the development of 1D short-range order.\cite{he05prb}
With lowering temperature, a rapid drop of $\chi$ at $T_N =$ 5.4 K indicates the onset of AF long-range order in zero magnetic field.
In the N\'eel state, the ordered moments of 2.18 $\mu_B$ per Co$^{2+}$ ion are aligned antiferromagnetically within the screw chains along the $c$-axis (right panel of Fig.~1).\cite{kawasaki11}

Quantum spin fluctuations strongly affect the N\'eel state, as evidenced by the field-induced transition from 3D AF long-range order to 1D quantum disorder.
The application of a magnetic field $H$ along the chain ($H\parallel c$) suppresses the N\'eel order and induces an incommensurably modulated AF phase above the critical field $H_c = 39$ kOe at low temperature.\cite{okunishi07,kimura08,kimura08a,suga08,yamaguchi11}
The suppression of the N\'eel state by the magnetic field can be understood by the appearance of a quantum critical state in the magnetic field, with the modulated AF structure above $H_c$ indicating the development of a longitudinal incommensurate spin correlation.

Thus, BaCo$_2$V$_2$O$_8$ possesses many interesting properties and these magnetic properties are worthwhile investigating microscopically, especially using NMR for its capability to explore local static and dynamical magnetic properties of quantum spin systems. 
Although Kuo has reported the absence of AF long-range order above 3.0 K at $H =$ 70.6 kOe $(>H_c)$ from the $^{51}$V-NMR spectrum,\cite{kuo09} there are no reports of NMR measurements in low magnetic fields where the N\'eel order survives.

In this paper, we report measurements of the $^{51}$V-NMR spectrum in the N\'eel state and of the nuclear spin-lattice relaxation rate in the paramagnetic state in low magnetic fields.
The former features a bi-rectangle spectrum at low temperature indicating two V sites subjected to different magnetic field magnitudes, which could be explained in terms of the classical and the pseudo-dipolar fields from Co$^{2+}$ spins. 
The incomplete bi-rectangle NMR spectrum between 3.5 K and $T_N=$ 5.4 K suggests a large spin fluctuation even in the N\'eel state, which can be related to a field-induced order-disorder transition.
The latter sheds light on the dynamics of the spin system in the paramagnetic state.
In such states, the linear relation between the nuclear spin-lattice relaxation rate $1/T_1T$ and the magnetic susceptibility $\chi$ indicates that the $Q$ component of magnetic susceptibility $\chi(Q)$ is roughly proportional to $\chi$.
The derivative of $1/T_1T$ with respect to $\chi$ around 150 K suggests a change in the AF spin-fluctuation spectrum.

\begin{figure}[tb]
\centering
\includegraphics[width=8.5cm]{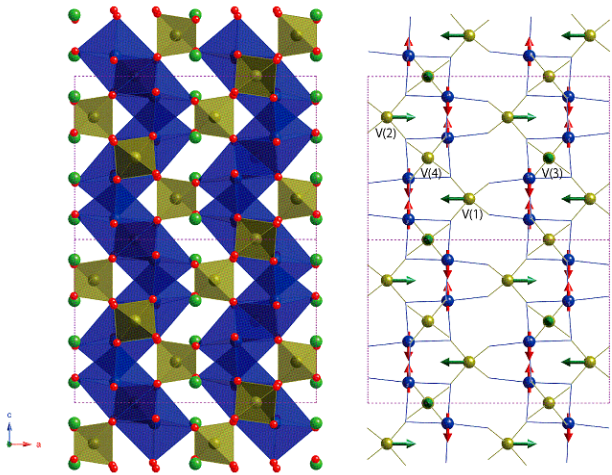}
\caption{(Color online) Left: Crystal structure of BaCo$_2$V$_2$O$_8$, where blue octahedra, gold tetrahedra, green large balls, and red small balls represent CoO$_6$, VO$_4$, Ba, and O, respectively. Right: Red and green arrows indicate the respective ordered magnetic moment of Co$^{2+}$ in the AF state\cite{kawasaki11} and the estimated dipolar field ${\vec H}_{\rm dip}$ at V sites (see text).}
\end{figure}

\section{Experiments}

Powder of BaCo$_2$V$_2$O$_8$ was synthesized by a standard solid-state reaction method using a stoichiometric mixture of BaCO$_3$ (4N), CoC$_2$O$_4$$\cdot$2H$_2$O (3N), and V$_2$O$_5$ (4N) as starting materials.
The synthesis of materials have been detailed elsewhere.\cite{he05cm}
X-ray diffraction and magnetic susceptibility confirm that the structural and magnetic properties obtained are consistent with those reported previously.\cite{he05prb}
The sample for the present NMR measurements is the same as that used in previous neutron diffraction and $\mu$SR experiments.\cite{kawasaki11}

The spin-echo signal of $^{51}$V nuclei in BaCo$_2$V$_2$O$_8$ was recorded by using a phase coherent pulsed NMR spectrometer.
To observe the spin-echo signal, we used a silver coil instead of the usual copper coil to avoid the $^{63}$Cu background signal, because the nuclear gyromagnetic ratio of $^{63}$Cu nucleus is very close to that of the $^{51}$V nucleus.
The $^{51}$V-NMR spectrum was obtained by tracing the intensity of the spin-echo signal as a function of the external magnetic field at the fixed frequency of 15.0 MHz\@.
The nuclear spin-lattice relaxation rate was measured by the saturation magnetization recovery method.

\section{Results and Discussion}

\subsection{$^{51}$V-NMR spectrum in the AF state}

\begin{figure}[tb]
\centering
\includegraphics[width=8.5cm]{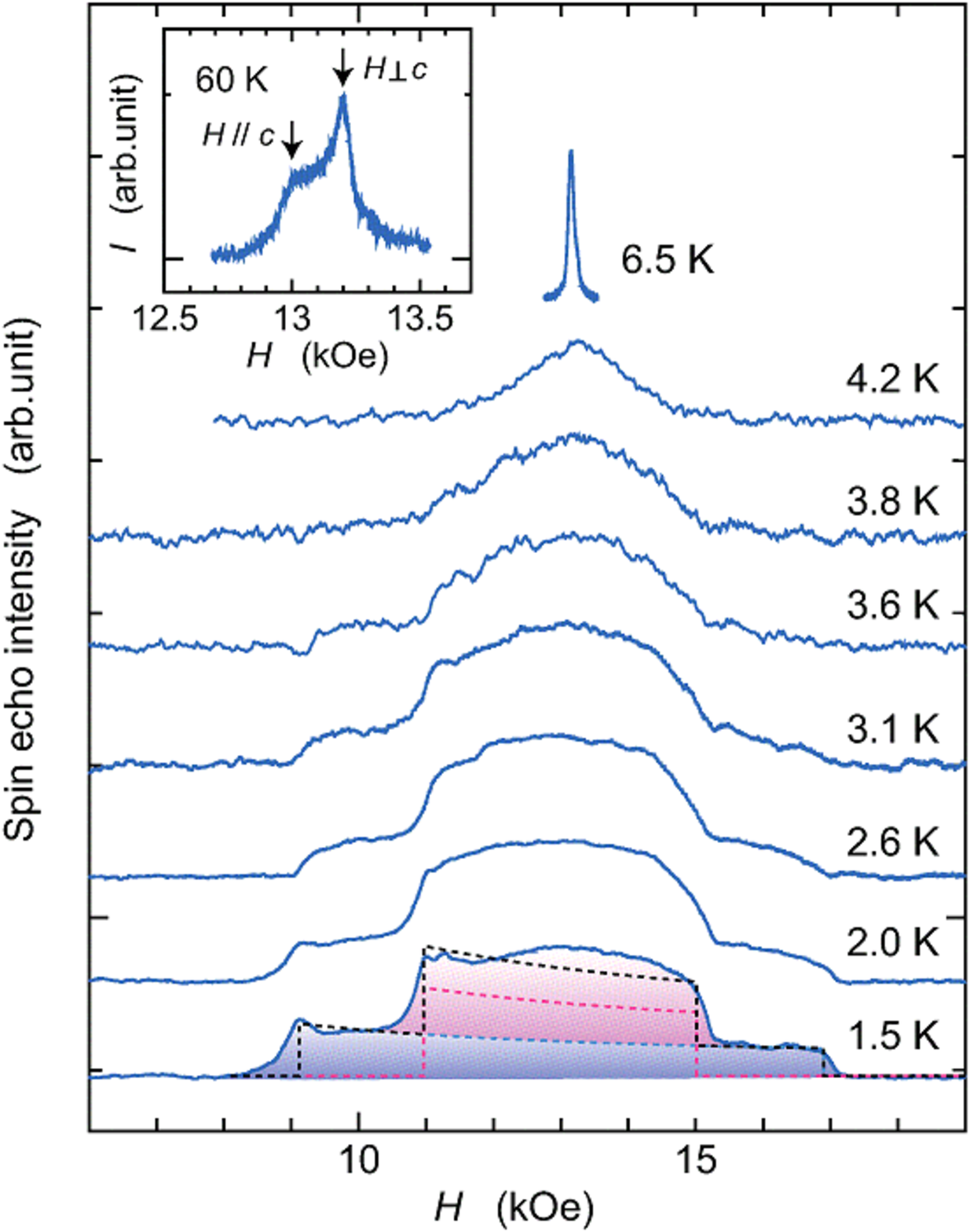}
\caption{(Color online) $^{51}$V-NMR spectra measured at 6.5 K and temperatures below $T_N=$ 5.4 K\@. The dashed lines are calculated from Eq.~(1) with $H_{\rm A1}=$ 2.1 and $H_{\rm A2}=$ 3.8 kOe. The inset shows the $^{51}$V-NMR spectrum observed at 60 K in the paramagnetic state.}
\end{figure}

\begin{table*}
\caption{Estimated dipolar field ${\vec H}_{\rm dip}$ at V sites arising from Co$^{2+}$ AF ordered moments in BaCo$_2$V$_2$O$_8$. The fractional coordinates of the V sites are expressed as: (1) $(0.08+0.34n+0.50m, 0.50n, 0.25)$,  (2) $(0.08+0.34n+0.50m, 0.50-0.50n, 0.75)$, (3) $(0.25+0.50n, 0.67+0.16m,  0.50|n-m|)$ and (4) $(0.25+0.50n, 0.17+0.16m,  0.50|n-m|)$, where $n$ and $m$ are 0 or 1.}
\medskip
\begin{center}
\begin{tabular*}{13.5cm}{llllll}
\hline\hline
site \ \  & ${\vec H}_{\rm dip}$ from 1st & ${\vec H}_{\rm dip}$ from 2nd & ${\vec H}_{\rm dip}$ from 3rd & ${\vec H}_{\rm dip}$ from 4th & Total ${\vec H}_{\rm dip}$\\
& nearest neighbors &nearest neighbors &nearest neighbors &nearest neighbors & \\
&(kOe) &(kOe) &(kOe) &(kOe) &(kOe) \\
\hline
V(1) &  $(-0.838, 0, 0)$ \ \ & $(0.347,0,0)$ &  $(0,0,0)$ \ \ & $ (-0.123,0,0)$ \ \ & $(-0.630,0,0)$ \\
  V(2)&   $  (0.838,0,0)$ & $ (-0.347,0,0)$ \ \ &  $ (0,0,0)$ &   $ (0.123,0,0)$&  $  ( 0.630,0,0)$  \\ 
  V(3) &    $(0,-0.844,0)  $&$(0,0.0587,0) $ & $ (0,0.159,0)$  &  $ (0,-0.183,0) $  & $(0,-0.779,0)$  \\
  V(4) &   $ (0,0.844,0)$   &$(0,-0.0587,0) $ & $(0, -0.159,0) $ & $(0,0.183,0)$ & $ (0, 0.779,0)$\\
\hline\hline
\end{tabular*}
\end{center}
\end{table*}

The $^{51}$V-NMR spectra measured at 6.5 K and at temperatures below $T_N=5.4$ K are shown in Fig.~2\@.
A broadening in the line width is observed below $T_N$, corresponding well to a magnetic phase transition.  
The peak position in the NMR spectrum at 4.2 K, which is not largely shifted from that at 6.5 K\@, indicates the AF nature of the magnetic phase transition.

Generally, a rigid AF spin arrangement produces a single rectangle NMR spectrum in powder samples (powder pattern).\cite{yamada86}
In powders, the angle between the direction of the external magnetic field ${\vec H}$ and that of internal magnetic field ${\vec H_A}$ due to the AF spins is randomly distributed.
As a result, the number of nuclei exposed to the resonant magnetic field $H_0$ with $H_0=\omega_0/\gamma=|{\vec H}+{\vec H_A}|$ in the external magnetic field between $H$ and $H+dH$, denoted by $f (H, H_A, H_0)dH$, is expressible as 
\begin{equation}
f(H,H_A,H_0)=\frac{1}{4H_A}\left|1+\frac{H_0^2-H_A^2}{H^2}\right|
\end{equation}
for $|H_0-H_A| \leq H\leq H_0+H_A$. 
Here, $\omega_0$ is the NMR frequency and $\gamma$ is the nuclear gyromagnetic ratio.
This equation gives a rectangle powder pattern.

In the present case, we observed a bi-rectangle powder pattern below around 3.5 K, indicating two V sites subject to two different values of the internal magnetic field.
We estimated the magnitudes of the internal field to be $H_{\rm A1}$ = 2.1 and $H_{\rm A2}$ = 3.8 kOe  from the corresponding rectangle lengths.  
The dashed lines superimposed on the lowest temperature spectrum in Fig.~2 are calculated from Eqs.~(1) with these $H_{\rm A1}$ and $H_{\rm A2}$ values,  as well as their convolution.
The near-equal areas of the two rectangles indicate that the numbers of $^{51}$V nuclei exposed to $H_{\rm A1}$ and $H_{\rm A2}$ are almost the same.
In addition, we measured the $^{51}$V-NMR spectrum at the lower frequency of 8.9 MHz (not shown) and confirmed that the directions of ${\vec H_A}$'s are not affected by the external magnetic field.

Slightly above 3.5 K, the bi-rectangle pattern becomes indistinct, most probably due to quantum spin fluctuations related to the field-induced order-disorder transition.
This is in clear contrast to the usual behavior in antiferromagnets, such as the Heisenberg system Cu$_2$(OH)$_3$Cl and the heavy-fermion compound CePd$_2$Si$_2$, where the rectangle powder pattern appears just below $T_N$.\cite{27, 28} 
We note that the rigid magnetic Bragg reflections have been observed just below $T_N$ in the neutron diffraction pattern of BaCo$_2$V$_2$O$_8$ in zero magnetic field.\cite{kawasaki11}
No indication of spin fluctuations in the neutron diffraction pattern, in contrast to the present NMR results, suggests either the quantum spin fluctuations dominate over the N\'eel order as the magnetic field approaches from below $H_c$ or the correlation time in the spin fluctuations is longer than the characteristic time of observation for thermal neutrons but shorter than that for NMR\@.

The internal magnetic fields could have two origins: the scalar-type transferred hyperfine field and the dipolar field from the Co$^{2+}$ magnetic moments.
The scalar-type transferred hyperfine interaction is written as $A{\vec I}\cdot{\vec S}$, where $A$ is the hyperfine coupling constant, ${\vec I}$ is the $^{51}$V nuclear spin and ${\vec S}$ is the Co$^{2+}$ electronic spin, yielding the transferred hyperfine field parallel or antiparallel to the Co$^{2+}$ magnetic moments.
In the crystal structure of BaCo$_2$V$_2$O$_8$, there are six nearest-, six second nearest- and six fourth nearest-neighboring Co$^{2+}$ spins for each V site.
These Co$^{2+}$ spins consist of three antiparallel pairs of spins with fractional coordinates $(x_1, y_1, z_1)$ and $(-x_1, y_1, -z_1)$, or $(x_1, y_1, z_1)$ and $(x_1, -y_1, -z_1)$.
The four third nearest-neighboring Co$^{2+}$ spins consist of two antiparallel pairs of spins.  
Thus, the scalar-type transferred hyperfine fields arising from the Co$^{2+}$ pairs of spins are expected to cancel out and are not responsible for the observed internal fields.

The other possible origin of the internal magnetic field is the dipolar interaction.
We estimated the dipolar field ${\vec H_{\rm dip}}$ at the V site arising from the Co$^{2+}$ spins in the magnetically ordered state of BaCo$_2$V$_2$O$_8$, where the total dipolar fields were obtained from the lattice sum over a large spherical domain centered on the V site.
The results of the estimation, summarized in Table I, indicate four V sites each with different magnitudes and directions of the total dipolar field.
In particular, the magnitudes of the total dipolar fields fall into two values: 0.630 kOe at the V(1) and V(2) sites and 0.779 kOe at V(3) and V(4) sites.
In contrast, the directions of total dipolar fields at the V(1) and V(2) sites align along the $a$-axis but opposed and at the V(3) and V(4) sites along $b$-axis and again opposed (see right panel of Fig.~1)\@.

These results are consistent with not only the observation of two V sites with different field magnitudes but also equal V site distributions as indicated by equal rectangular areas in the powder pattern.  
In addition, the ratio between magnitudes of estimated dipolar fields, $0.779/0.630=1.24$, is comparable to the ratio $H_{\rm A2}/H_{\rm A1}=3.8/2.1=1.81$.
These agreements indicate that the dipolar field dominates the internal field at the V site.
Support also comes from our preliminary $^{51}$V-NMR measurements on single crystalline BaCo$_2$V$_2$O$_8$ below $T_N$\@, from which the remarkable split in the resonance lines for $H\parallel a$ and $H\parallel b$ indicate that the internal magnetic fields align along the $a$-axis at the V(1) and V(2) sites and along the $b$-axis at the V(3) and V(4)\@.

The estimated magnitudes of the dipolar fields, 0.630 and 0.779 kOe, however, are about a quarter of the observed values $H_{\rm A1}$ = 2.1 and $H_{\rm A2}$ = 3.8 kOe.
This difference could be explained by considering the pseudo-dipolar field.
The pseudo-dipolar interaction between nuclear spins has been proposed by Bloembergen and Rowland for Tl nuclei in Tl metal and Tl$_2$O$_3$.\cite{29}
In Co dilute alloys, Portis and Kanamori have proposed a pseudo-dipolar field on Co and Fe nuclei from the neighboring electron spins that arises from the distortion of the neighboring $d$ shell.\cite{portis62}
The pseudo-dipolar field has the same angular dependence as the classical dipolar field, although the distance dependencies are different, {\it i.e.}, the pseudo-dipolar interaction is effective at shorter range compared with the classical dipolar interaction.
The pseudo-dipolar fields are estimated to be about 30 times larger in Tl metal but similar in Tl$_2$O$_3$ as large as the classical dipolar interaction,\cite{29} whereas the pseudo-dipolar field on Co in Fe, Fe in Co, and Co in Ni are estimated to be respectively 2, 1 , and 5 times larger.\cite{portis62}
In BaCo$_2$V$_2$O$_8$, the present experimental results could be explained in terms of the classical and the pseudo-dipolar fields, if we assume that the pseudo-dipolar field on V from Co electron spins is approximately three times larger than the classical dipolar field and that the dipolar and pseudo-dipolar fields point to the same direction.

\subsection{Spin dynamics in the paramagnetic state}

In the $^{51}$V-NMR powder spectrum observed at 60 K in the paramagnetic state (see inset of Fig.\ 2), the large peak and shoulder are associated with $H\perp c$ and $H\parallel c$, respectively.
We measured the spin-lattice relaxation rate $1/T_1$ for $H\perp c$ at the magnetic field where the spin-echo intensity has a maximum.

We calculated the temperature dependence of $1/T_1$ divided by temperature, $1/T_1T$ (Fig.~3).
With decreasing temperature, starting from 300 K, $1/T_1T$ gradually increases.
The rate of change becomes larger below around 100 K, followed by a steep up-turn below 14 K\@.
The increase in $1/T_1T$ below around 100 K indicates the development of 1D short-range order among Co$^{2+}$ spins.
The steep upturn in $1/T_1T$ below 14 K indicates a typical critical behavior near $T_N$ as the dynamics of the Co$^{2+}$ spins are gradually slowing down.
A similar temperature dependence in the relaxation rate is also observed from $\mu$SR measurements in zero magnetic field.\cite{kawasaki11,mansson12}

\begin{figure}[tb]
\centering
\includegraphics[width=8.2cm]{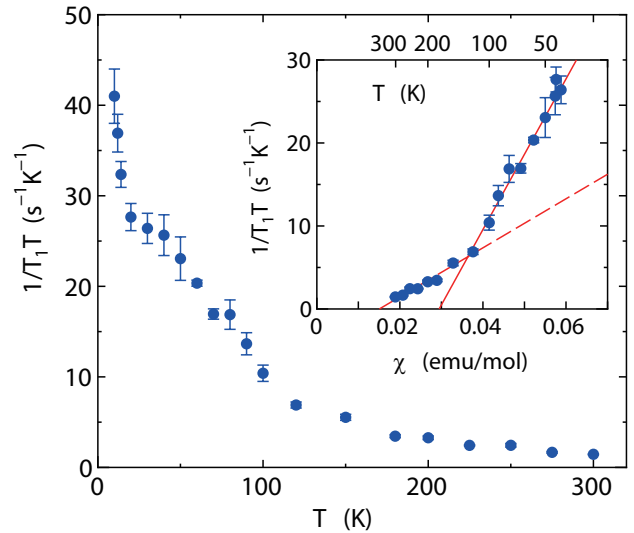}
\caption{(Color online) The temperature dependence of $1/T_1T$. The inset shows $1/T_1T$ against the magnetic susceptibility $\chi$ with the temperature as an implicit parameter. The solid and dashed lines show the linear fits between $1/T_1T$ and $\chi$ for 20 K $<T<$ 150 K and 150 K $<T<$ 300 K, respectively.}
\end{figure}

In general, $1/T_1T$ is given in terms of the imaginary part of the $\bm{q}$-dependent dynamical susceptibility $\chi(\bm{q}, \omega)$ and expressed by
\begin{equation}
\frac{1}{T_1T}=\frac{\gamma^2k_B}{2\mu_B^2}\sum_{\bm{q}} A(\bm{q})^2\frac{\chi''(\bm{q},\omega)}{\omega}, 
\end{equation}
where $k_B$ is the Boltzmann constant and $A(\bm{q})$ the $\bm{q}$-dependent hyperfine coupling constant.\cite{moriya63}
The form factor $F(\bm{q})$ is defined by $A(\bm{q})^2$\@.
 For $\chi(\bm{q},\omega)$, a Lorentzian spectrum having a $\bm{q}$-dependent characteristic energy $\Gamma_{\bm{q}}$ of the AF spin fluctuations is assumed to have the form
\begin{equation}
\chi(\bm{q},\omega)=\frac{\chi(\bm{Q})}{1+(\bm{q}-\bm{Q})^2\xi^2-i\hbar\omega/\Gamma_{\bm{q}}},
\end{equation}
where $\xi$ is the correlation length and $\bm{Q}$ the AF wave number vector.
Hereafter, we approximate this spin system as 1D one.
$q$ is the wave number for the chain direction and $Q=\pi/a$ is the AF wave number with the distance $a$ between spins.

The AF spin fluctuations of Co$^{2+}$ can produce fluctuations in the internal magnetic field at the V sites through the scalar-type transferred hyperfine interaction and the dipolar interaction.
As discussed previously, the scalar-type transferred hyperfine fields are expected to cancel out at the V sites in the N\'eel state, {\it i.e.}, the corresponding form factor $F(q)$ is nearly zero at $q=Q$.
Even if $\chi''(q,\omega)$ has a sharp peak at the AF wave number $Q = \pi/a$, the $Q$ component of $A(q)^2\chi''(q, \omega)$ in Eq.~(2) is zero and does not contribute to $1/T_1T$\@.
We calculated the form factor $F(q)$ for V in BaCo$_2$V$_2$O$_8$ due to the scalar-type transferred hyperfine field from the six nearest-neighboring Co ions.
The calculation gives $F(q)=[4A\cos(1.0469q)+2A\cos(3.1406q)]^2$, where $A$ is the transferred hyperfine field per nearest-neighboring Co ion.
$F(q)$ is nearly zero in the range of $0.6Q<q<Q$ and has a maximum at $q=0$ (see Fig.\ 4).
This $q$ dependence of $F(q)$ means that the $q$ components from $q=0$ to $0.6Q$ contribute to $1/T_1T$ through the scalar-type transferred hyperfine field.

\begin{figure}[tb]
\centering
\includegraphics[width=8.2cm]{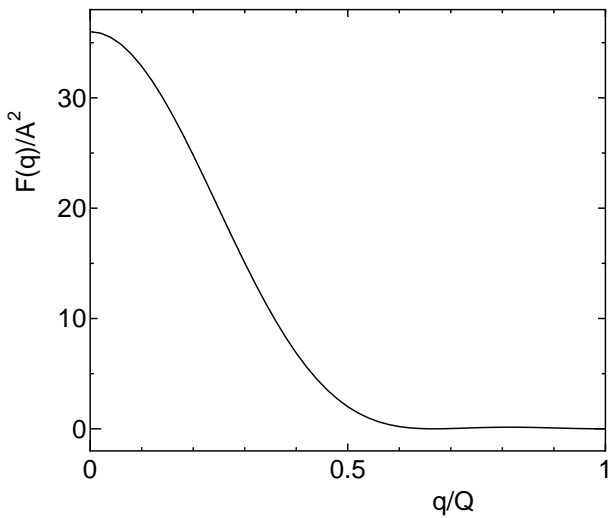}
\caption{The $q$ dependence of form factor $F(q)$ for V due to the scalar-type transferred hyperfine field from the six nearest-neighboring Co ions.}
\end{figure}

In contrast, the dipolar field produces an internal magnetic field at V site in the N\'eel state, {\it i.e.}, the form factor $F(q)$ due to the dipolar field is not zero at $q=Q$.
If $\chi''(q, \omega)$ has a sharp peak at $Q = \pi/a$, $1/T_1T$ from fluctuations of the dipolar field $(1/T_1T)_{\rm dip}$ is
\begin{equation}
\left(\frac{1}{T_1T}\right)_{\rm dip}\propto\chi''(Q)\propto \frac{\pi\chi(Q)}{\Gamma_Q}
\end{equation}
from Eqs.\ (2) and (3).
In BaCo$_2$V$_2$O$_8$, the observed $1/T_1T$ is considered to be dominated by the $q=Q$ component of AF spin fluctuations through the dipolar-type coupling between V and surrounding Co$^{2+}$ ions, because generally $\chi(q)$ has a large peak at $q=Q$ in AF systems.

We show $1/T_1T$ against $\chi$ (see inset of Fig.~3) with temperature as an implicit parameter.
The magnetic susceptibility $\chi$ for powder are referred from the literature.\cite{he05prb}
The linear fits between $1/T_1T$ and $\chi$ are found in the temperature ranges 20-150 K and 150-300 K, respectively.
Similar linearity has been widely observed in other quasi-1D spin systems\cite{ghoshray05,shimizu95,izumi03,ishida96}, and this fact indicates that $\chi(Q)$ is roughly proportional to $\chi$ in these systems as expected from Eq.~(4)\@.
For BaCo$_2$V$_2$O$_8$, therefore, $\chi(Q)$ is roughly proportional to $\chi$ in the respective temperature regions and both $1/T_1T$ and $\chi$ originate from the same Co$^{2+}$ spins.
A change in slope for $1/T_1T$ against $\chi$ around 150 K\@ is observed in Fig.\ 3.
Moreover, the hyperfine coupling constant between $^{51}$V nuclei and Co$^{2+}$ spins is independent of temperature as indicated by the $K$-$\chi$ plot in single crystalline BaCo$_2$V$_2$O$_8$.\cite{kuo09}
These results suggest a change in the AF spin-fluctuation spectrum, for example, the change in $\Gamma_{\bm{q}}$ and/or the change in $\xi$, at around 150 K\@.

\section{Conclusion}
We have investigated the static and dynamic magnetic properties in the quasi-one dimensional antiferromagnet BaCo$_2$V$_2$O$_8$ by $^{51}$V NMR\@. 
The well-defined bi-rectangle powder pattern at low temperature indicates two V sites subjected to two different internal magnetic field magnitudes, $H_{A1}=2.1$ and $H_{A2}=3.8$ kOe, although V atoms occupy only one magnetically symmetric site where the scalar-type transferred field from the surrounding Co$^{2+}$ spins is zero. 
These internal fields could be explained by taking account of the classical and the pseudo-dipolar fields from Co$^{2+}$ spins. 
The indistinct bi-rectangle powder pattern just below $T_N$ indicates a large spin fluctuation even in the N\'eel state, which could be related to the field-induced order-disorder transition at $H_c=39$ kOe.
In the paramagnetic state, the nuclear spin relaxation is dominated by AF spin fluctuations through the dipolar-type coupling between V and surrounding Co$^{2+}$ ions.
Linearity between $1/T_1T$ and $\chi$ indicates that $\chi(Q)$ is roughly proportional to $\chi$, but the change in the AF spin-fluctuation spectrum appears to occur around 150 K\@.

\end{document}